\documentclass[aps,prb,showpacs,twocolumn,floats]{revtex4}
\usepackage[bookmarks=false]{hyperref}
\usepackage{amssymb}
\usepackage{amsbsy}
\usepackage{amsmath}
\usepackage{graphicx}
\usepackage{times}
\usepackage{color}
\usepackage{setspace}
\usepackage{bm}
\newcommand\bea{\begin{eqnarray}}
\newcommand\eea{\end{eqnarray}}
\newcommand\beq{\begin{equation}}
\newcommand\eeq{\end{equation}}

\begin{document}

\title{Spin Transport and Spin Pump in Graphene-like Materials: Effect of tilt in Dirac cones}

\author{Debabrata Sinha}

\affiliation{Theoretical Physics Department, Indian Association for
the Cultivation of Science, Jadavpur, Kolkata-700032, India.}

\date{\today}

\begin{abstract}
We study the spin transport phenomena in two-dimensional graphene-like materials with arbitrary tilted Dirac cones. The tilt arises due to next-nearest hopping when the bottom of the conduction band and top of the valence band does not simultaneously coincide at Dirac point. We consider normal-ferromagnetic-normal (N-F-N) junction of the materials and using the generalized scattering approach calculate the spin current. Here, we show that tilting the Dirac cones can strongly change the transport properties by modifying the period of oscillation of the spin current. The spin conductance can be effectively tuned by the tilt with taking advantage of the modified interference condition. A pure spin current reversal also possible with a smooth variation of the tilting. We further study the spin current by the adiabatic precession of a doped ferromagnet on top of the material. It is shown that the spin-mixing conductance and hence the spin current can become zero by turning the tilt of the Dirac cone. Our findings provide an efficient way towards high controllability of spin transport by tuning the tilt of the ferromagnetic junction and can be very useful in the field of spintronics. The model also presents a simplified way to measure the tilt of Dirac cone of those materials.
\end{abstract}
\maketitle

\section{Introduction}
The generation and control of pure spin current are the most challenging issue in the field of spintronics\cite{Hanson-RMP-07, Garcia-Sci-10, Zutic-Sci-12}. The essential condition for the spin current transport is the net transfer of spin angular momentum which occurs due to nonequilibrium imbalance between the spin density of the two spin components. In the last few years, plenty of models have been proposed and implemented to control the spin current in ferromagnetic graphene, $e.g.,$ by gate voltage\cite{Yoko-08,Take-PRB 11}, strain \cite{Zein-JMMM 17,Peet-PRL 14,Feng-APL 11,Qing-Sci 16}, optical irradiation \cite{Moha-Super 16,Le-JPD 16} etc. Although the neutral graphene is originally non-magnetic, the proximity effect permits it to be ferromagnetic. The exchange coupling in it can be tuned by the in-plane external electric field \cite{Son-Nat 06}and thus has huge applications in controlling the spin transport.

Spin injection into various nonmagnetic devices and spin pumping by time-dependent external parameters also constitute another important avenue of spintronics. In recent years,  several studies have been done both theoretically and experimentally on the electrical spin injection in topological insulator, semiconductors\cite{Matt-Sci 16,Krish-PRB 12,Ohno-Nat 99,Rashba-PRB 00,Jedema,Sarma}, spin hall effect \cite{Niu,Kato,Wunderlich}, adiabatic spin pumping via time-dependent gate voltage\cite{Zhang-APL 11}, etc. However, it is shown that the resulting spin conductance is relatively small due to conductance mismatch problem and thus requires high-quality tunnel barrier to overcome the issue\cite{Tang-PRB 13,Luis-APL 11}. It is also possible to generate pure spin current by pumping via a ferromagnetic junction with rotating magnetization\cite{Mikha-PRB-10,Rah-JAPPL 15,Ino-PRB 16,Tse-PRB 02,Lenz-PRB 04}. The spin current generated by the adiabatic precession of magnetization vector in the ferromagnetic region does not suffer conductance mismatch problem and it is the source of many physical phenomena $e.g.$, the spin Seebeck effect \cite{Uchida-Nat 10}. Another way to see the one-way spin current was proposed in Ref.(\cite{Luis-PRB-16}) very recently.

Since the discovery, graphene has drawn much attention in condensed matter physics owing to its rich potential applications. Grahene has hexagonal lattice structure made of carbon atoms and its low energy spectrum is described by pseudo-relativistic Dirac equation. In the first Brillouin zone (BZ), the eigenenergy has linear dispersion relation around two inequivalent points $K$ and $K'$, known as Dirac points. In contrast, there are a wider class of Dirac materials where the low energy effective theory is described by tilted Dirac cones as in the case for organic materials $\alpha(BEDTTTF)_2I_3$\cite{Goe-PRB 08}, borophene\cite{Zabo-PRB 16},\cite{Ale-PRB 16} and in a certain class of topological insulators \cite{Vary-PRB 17}. The Dirac cones of these materials are seen to be strongly anisotropic and tilted in wave-vector energy space. The tilted Dirac cone also appear in graphene by a quinoid-type lattice deformation\cite{Goe-PRB 08, Pauling}. The recent development of cold atom in an optical lattice allows one to deform the honeycomb lattice with help of laser intensity and wavelength to achieve the tilted Dirac cones\cite{Pyri-PRL 17,Shi-PRL 07}.

The tilt of the Dirac cone is usually neglected since it does not affect the topology of the band structure. However, tilt in the Dirac cone shows many fascinating behaviour in particular in quantum transport\cite{Viet},\cite{Tre-PRB 15}, optical conductivity \cite{Car-PRB 16}, magnetoplasmons \cite{Sara-PRB 14} and ferromagnetic spin polarization \cite{Hira-Nature 16}. Here, we demonstrate that the tilt leads to clear signature of quantum spin transport and it can stand as an efficient tuning parameter other than doping, strain or optical irradiation. Particularly, we have shown tilt modify the quantum interference condition in normal/ferromagnet/normal junction resulting in a spin current reversal.
Moreover, we reveal that the spin current pumped into two adjacent reservoirs can be controllable by tilt which leads to the spin valve effect.

\section{Spin Transport and Effect of tilt}
\begin{figure}
\center
\includegraphics[width=0.15\textwidth]{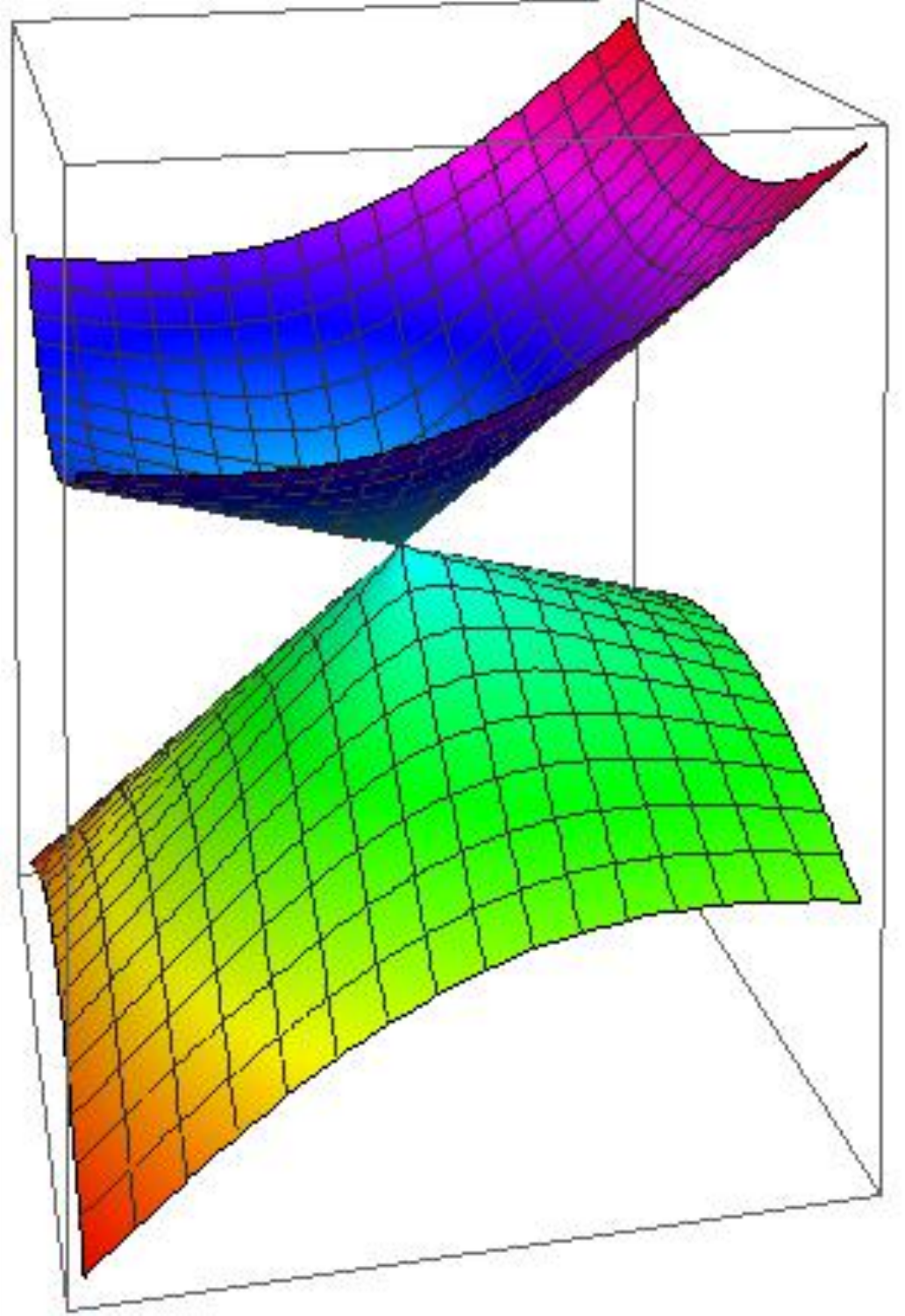}
\caption{Tilted Dirac cone $v_t<1$.}
\label{tilt-dirac}
\end{figure}

In the vicinity of a nodal point, a generic Dirac Hamiltonian with finite tilt along $x-$direction without anisotropy can be expressed as\cite{Tre-PRB 15},
\begin{eqnarray}
\mathcal{H}=\hbar v_{1}k_x  \sigma_0 +\hbar v_{2}(k_x \sigma_x +k_y \sigma_y)
\label{simplify-hamil}
\end{eqnarray}
The eigenvalues of the Hamiltonian Eq.(\ref{simplify-hamil}) lie on the hyperboloid sheets $E_\pm$,
\begin{eqnarray}
E_{\pm}=\hbar v_1 k_x\pm \hbar v_2 \sqrt{k^2_x+k^2_y}
\end{eqnarray} 
The "tilt" term is defined in the rest of the paper as $v_t=v_1/v_2$. It is straightforward to show that the eigenfunctions of Eq.(\ref{simplify-hamil}) don't depend on the tilt. So, it leaves the Berry curvature unchanged and hence the topology of band structure. However, the dependency of eigenenergy, namely, the Fermi level on tilt plays the important role in quantum transport. For $v_t=0$, there is no tilt and we get back the Dirac Hamiltonian for graphene-like materials. For finite value of $v_t$, the Dirac cone has been tilted along the $x$-direction as illustrated in Fig.(\ref{tilt-dirac}). Although, $v_t$ can take an arbitrary value in three dimensional Dirac materials e.g., Weyl semimetals \cite{Ale-Nature 15}but in two dimensions the value of $v_t$ is seen to be less than one. Two types of fermions ($v_t<1$ and $v_t>1$) have attracted much attention due to their unique physical properties\cite{Been-PRL 16,Hou-PRB 17}. However, here we concentrate tilt Dirac cone with $v_t<1$ and leave the case $v_t>1$ for future studies.

Here we theoretically investigated spin transport in two-dimensional normal/ferromagnetic/normal junction of graphene-like materials. We consider the tilt Dirac cone is only in the ferromagnetic region. The ferromagnetic part of these materials can be realized via proximity effect by attaching a magnetic insulator close to them. The estimated induced exchange field in order of few meV. A gate voltage is attached to the ferromagnetic region to tune the Fermi level compare to the normal regions. The junction interfaces are parallel to the $y$-axis and located at $x=0$ and $x=D$. Since there is valley degeneracy, we focus only on the Hamiltonian at the $K$ point. The Hamiltonian is given by,

\begin{eqnarray}
\mathcal{H}=\hbar v_1 k_x \sigma_0 +\hbar v_2 (k_x\sigma_x+k_y\sigma_y)-V(x)
\end{eqnarray}
where $V(x)=E_{F}$ in normal region and $V(x)=E_{F}+U+\sigma J$ (with $\sigma=\pm 1$) in the ferromagnetic region. We further consider that the term $v_1$ is finite in the ferromagnetic region ($0<x<D$) and zero otherwise. We choose the velocity $v_2 \simeq 10^6$ m/s i.e. same as Fermi velocity of graphene. $E_{F}$ is the Fermi energy and $U$ is the gate voltage responsible for Fermi energy shift in the ferromagnetic region, $J$ is the exchange field. The wavefunction in $x<0$ region is given by,
\begin{eqnarray}
\Psi_{I}&=&\begin{pmatrix} 1 \\ \gamma e^{i\phi} \end{pmatrix} e^{ik_x x}+ r^{\sigma}_{k}\begin{pmatrix} 1 \\ -\gamma e^{-i\phi} \end{pmatrix} e^{-ik_x x}
\end{eqnarray}
in $0<x<D$ is reads as,
\begin{eqnarray}
\Psi_{II}&=&a^{\sigma}_{k}\begin{pmatrix} 1 \\ \gamma^{\sigma}_1 \mathcal{P}_1 \end{pmatrix} e^{iq^{+}_x x}+ b^{\sigma}_{k}\begin{pmatrix} 1 \\ \gamma^{\sigma}_1 \mathcal{P}_2 \end{pmatrix} e^{iq^{-}_x x}
\end{eqnarray}
and for $x>D$ is
\begin{eqnarray}
\Psi_{III}&=& t^{\sigma}_{k}\begin{pmatrix} 1 \\ \gamma e^{i\phi} \end{pmatrix} e^{ik_x x}
\end{eqnarray}
where $\phi$ denotes the angle of incidence with respect to the barrier, $k_x=(|E_F|/\hbar v_{F})\cos \phi, k_y=(|E_F|/\hbar v_{F})\sin \phi$ and $\gamma=Sign(E)$. $\gamma^{\sigma}_{1}=Sign[(E+E_F+U+\sigma J)]$, $\mathcal{P}_{1,2}=(q_{x+,-}+iq_y)/\sqrt{q^2_{x,+,-}+q^2_y}$ where the expression of $q_{x\pm}$ is given,
\begin{eqnarray}
q^{\pm}_x=\frac{1}{\hbar v_2}[\alpha_1\mp \alpha_2]
\label{sol}
\end{eqnarray}
and
\begin{eqnarray}
\alpha_1&=&\frac{v_t(E+E_F+U+\sigma J)}{(v^2_t-1)}\nonumber\\
\alpha_2&=&\frac{\sqrt{(E+E_F+U+\sigma J)^2+(v^2_t-1)\hbar^2v^2_2q^2_y}}{(v^2_t-1)}
\label{sol-part}
\end{eqnarray}
Note that, we get back the solutions for normal graphene in the limit $v_t=0$\cite{Yoko-08}. Because of the translational symmetry in the $y$-direction, the momentum parallel to the $y$-axis is conserved. Now matching the wave function at the interface at $x=0$ and $x=D$, we obtain the transmission coefficient given by,

\begin{eqnarray}
t_{\sigma}(\Phi)=\frac{\cos \phi (\mathcal{P}_1-\mathcal{P}_2)Exp[i\frac{\alpha_1 D}{\hbar v_2}]}{X_\sigma}
\label{trans}
\end{eqnarray}
where,
\begin{eqnarray}
X_{\sigma}&=&(\mathcal{P}_1-\mathcal{P}_2)\cos\phi \cos\theta-(\mathcal{P}_1+\mathcal{P}_2)\sin\phi \sin\theta\nonumber\\&&+i\gamma \gamma^{\sigma}_1\sin\theta(1-\mathcal{P}_1\mathcal{P}_2)
\end{eqnarray}
with $\theta=\alpha_2 D/\hbar v_2$. Then the dimensionless spin-resolved conductances $G^{\sigma}$ are given by,
\begin{eqnarray}
G_{\sigma}=\frac{1}{2}\int^{\pi/2}_{-\pi/2} d\phi \cos\phi |t_{\sigma}(\phi)|^2
\end{eqnarray}
The spin conductance is defined as $G_{s}=G_{\sigma=+1}-G_{\sigma=-1}$. Here we focus on the zero-bias condition i.e., $E=0$. We define here $\chi_1=UD/\hbar v_2$, $\chi_2=JD/\hbar v_2$ and $k_f=E_{F}/\hbar v_2$. In the limit $U+\sigma J>>E_F$, we can write $\theta =\chi_{\sigma}=\chi_1+\sigma \chi_2$ for $v_t=0$ and the transmission coefficient Eq.(\ref{trans}) simplifies in absence of tilt is given by \cite{Yoko-08},
\begin{eqnarray}
t_{\sigma} \simeq \frac{\cos \phi e^{-ik_xD}}{\cos\chi_{\sigma}\cos\phi -i\gamma \gamma^{\sigma}_1\sin \chi_\sigma}
\label{trans-nor}
\end{eqnarray}

\begin{figure}
\center
\includegraphics[width=1.4in]{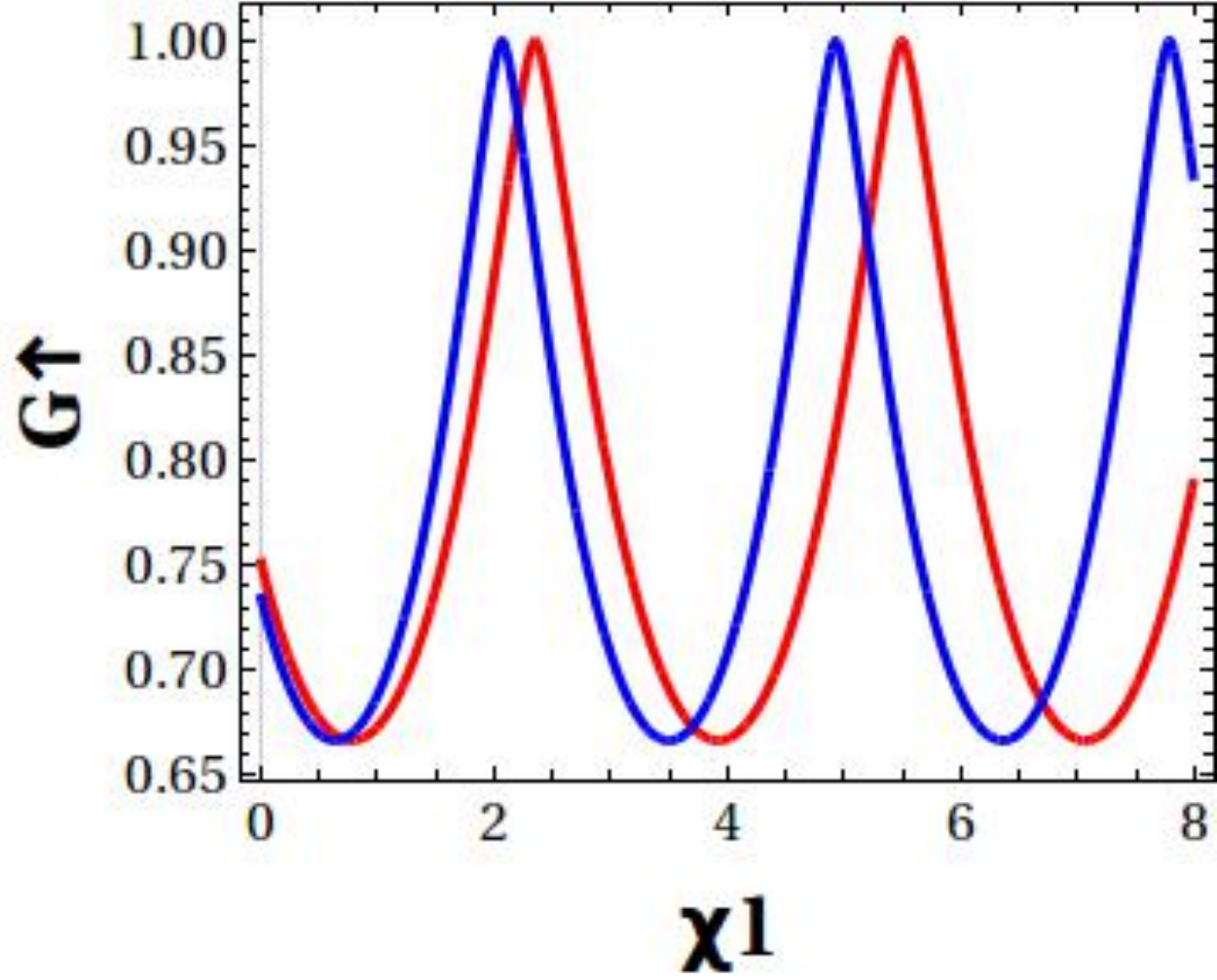}
\includegraphics[width=1.73in]{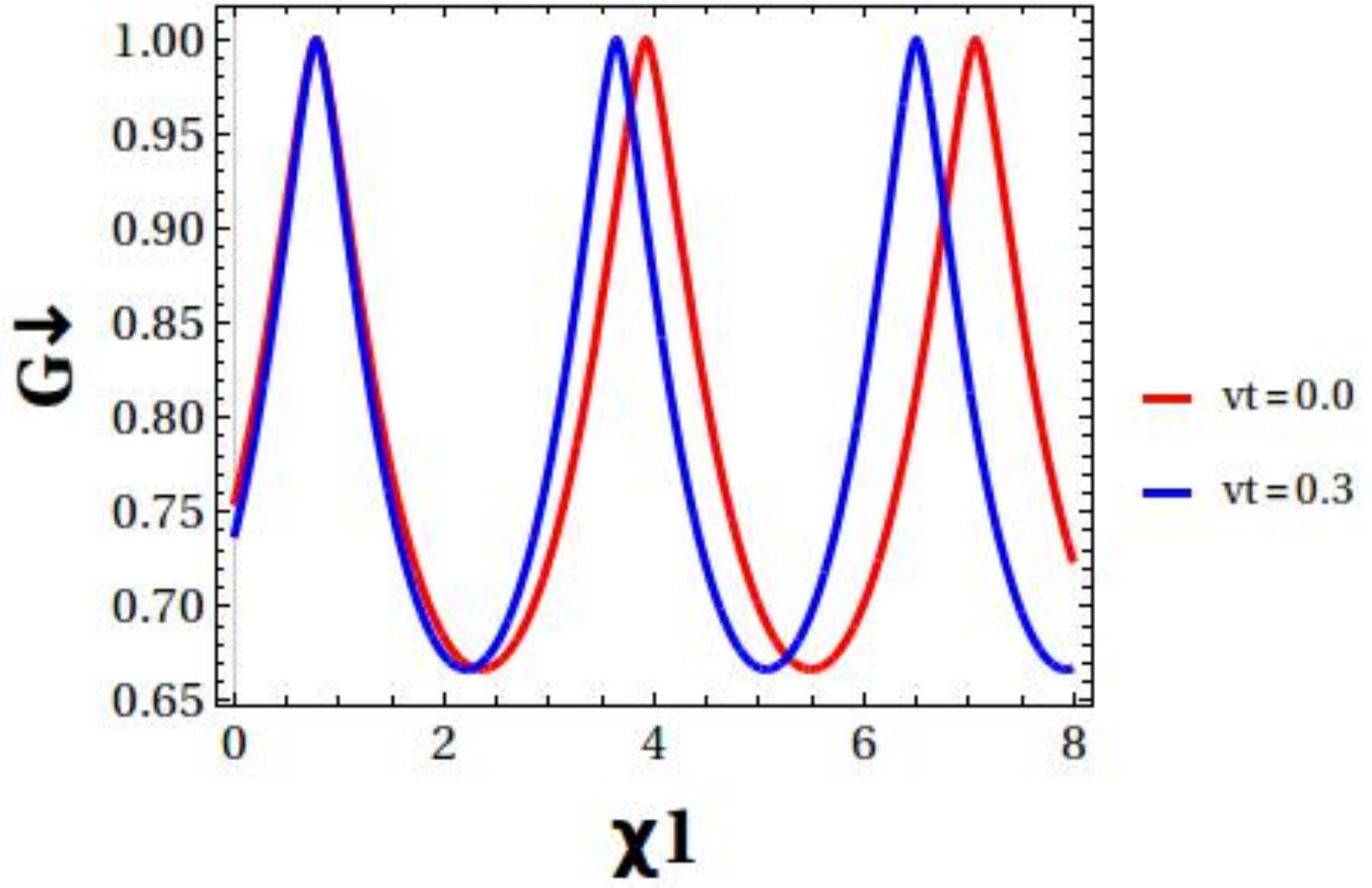}
\caption{Spin conductance ($G_{\uparrow}$, $G_{\downarrow}$) as a function of $\chi_1$ for two different values of $v_t$. The other parameters are $\chi_2=\pi/4$ and $k_{F}D=0$.}
\label{spin-cond-ud}
\end{figure}
\begin{figure}
\center
\includegraphics[width=1.4in]{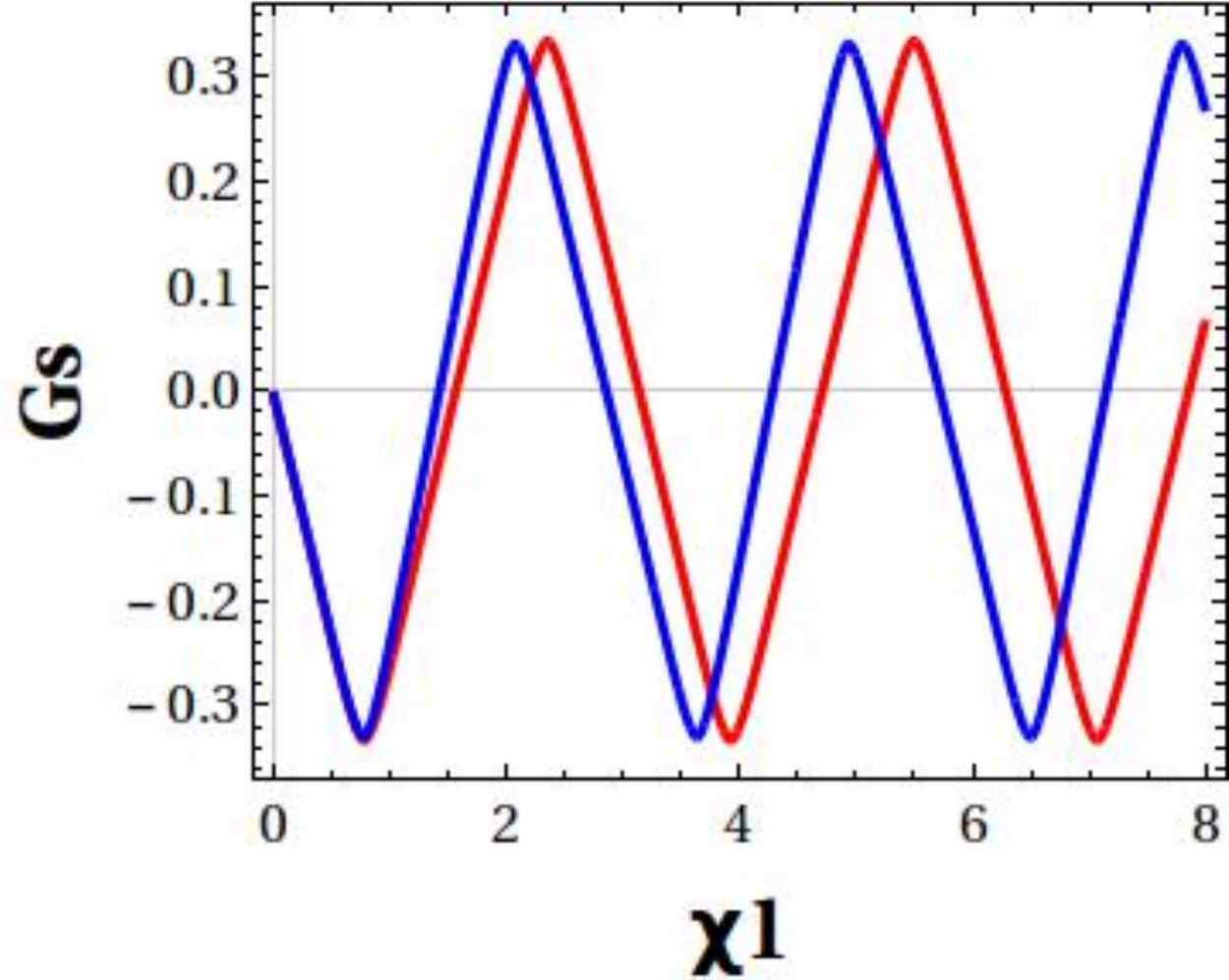}
\includegraphics[width=1.73in]{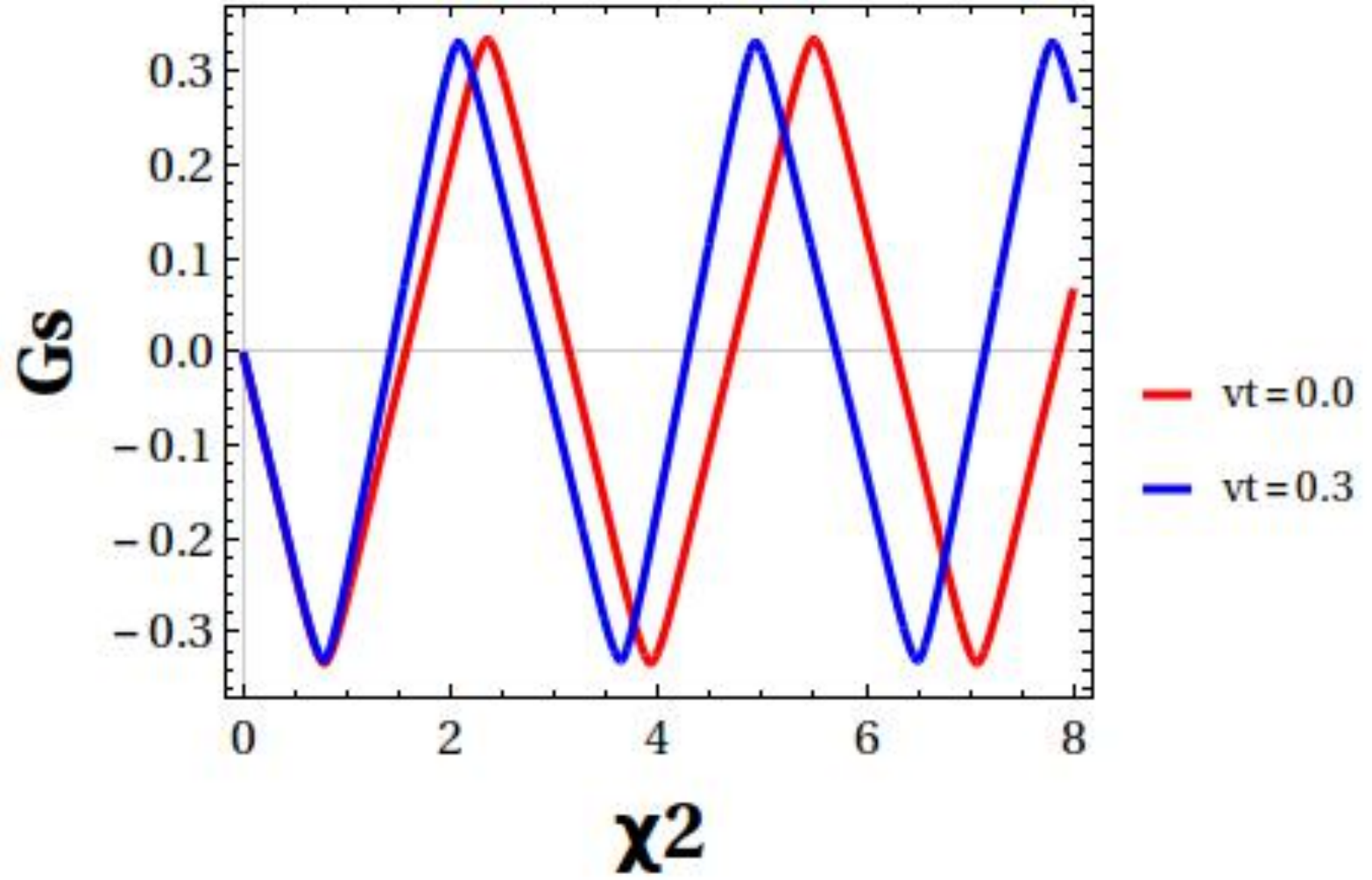}
\caption{Spin conductance ($G_s$) as a function of $\chi_1$ and $\chi_2$ for two different values of $v_t$. We fix $\chi_2=\pi/4$ and $\chi_1=\pi/4$ respectively. The value of $k_{F}D$ is zero.}
\label{spin-cond}
\end{figure}
We first discuss spin transport in the absence of tilt in the Dirac cone. From Eq.(\ref{trans-nor}), it is clear that the transmission coefficient $T_{\sigma}$ reaches its maximum (unity) at $\chi_\sigma=n\pi$ (since the Fermi level resides close to the Dirac point of one spin subband and maximize the transport of other spin component and vice versa) and becomes minimum at $\chi_{\sigma}=(2n+1)\pi/2$. We expect $\pi$ periodicities of spin conductances $G_{\uparrow}$ and $G_{\downarrow}$ with $\chi_{\sigma}$ or gate voltage $\chi_1$ as seen in Fig.(\ref{spin-cond-ud}). The spin conductances $G_{\uparrow,\downarrow}$ also show maximum and minimum values of 1 and $2/3$ respectively at $\chi_{\uparrow,\downarrow}=n\pi$ and $(2n+1)\pi/2$. The phase difference between the two spin conductances $G_{\uparrow,\downarrow}$ is $\chi_{\uparrow}-\chi_{\downarrow}=2\chi_2$. The spin conductance $G_s$ become large in magnitude ($\pm 1/3$) at those values of $\chi_1$ for which $G_{\uparrow}$ and $G_{\downarrow}$ become maximum and minimum respectively or vice versa. Such events only occur when the following condition is satisfied: $2\chi_2=(2n+1)\pi/2$ (i.e., $J/E_F=(2n+1)\pi/4k_{F}D$). We choose the parameter $\chi_2=\pi/4$ in the left part of Fig.(\ref{spin-cond}) and it is seen that the spin conductance $G_s$ has a maximum and minimum value of $1/3$ and $-1/3$ and a $\pi$ periodicity with $\chi_1$ \cite{Yoko-08}. A similar behavior of spin conductance with proximity strength ($\chi_2$) also seen in the right part of Fig.(\ref{spin-cond}). However, for large value of $k_F$ ($E_F \gtrsim (U\pm J)$), the $\pi$ periodicity is broken (not shown). Note that, since $\chi_1=k_F U D/E_F$, the $\pi$ periodicity can be recovered again for large value of $D$.

\begin{figure}
\center
\includegraphics[width=2.5in]{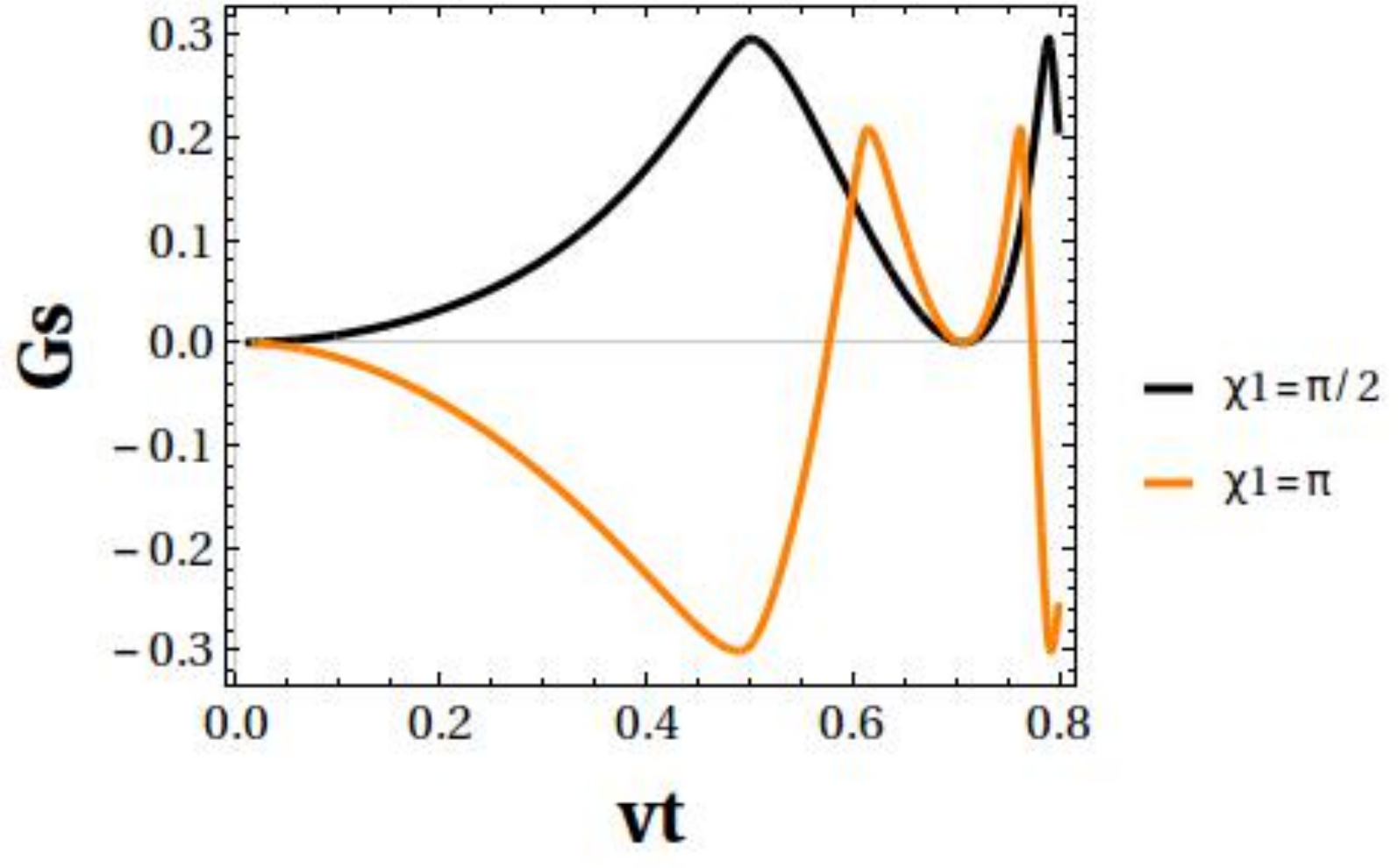}
\includegraphics[width=2.5in]{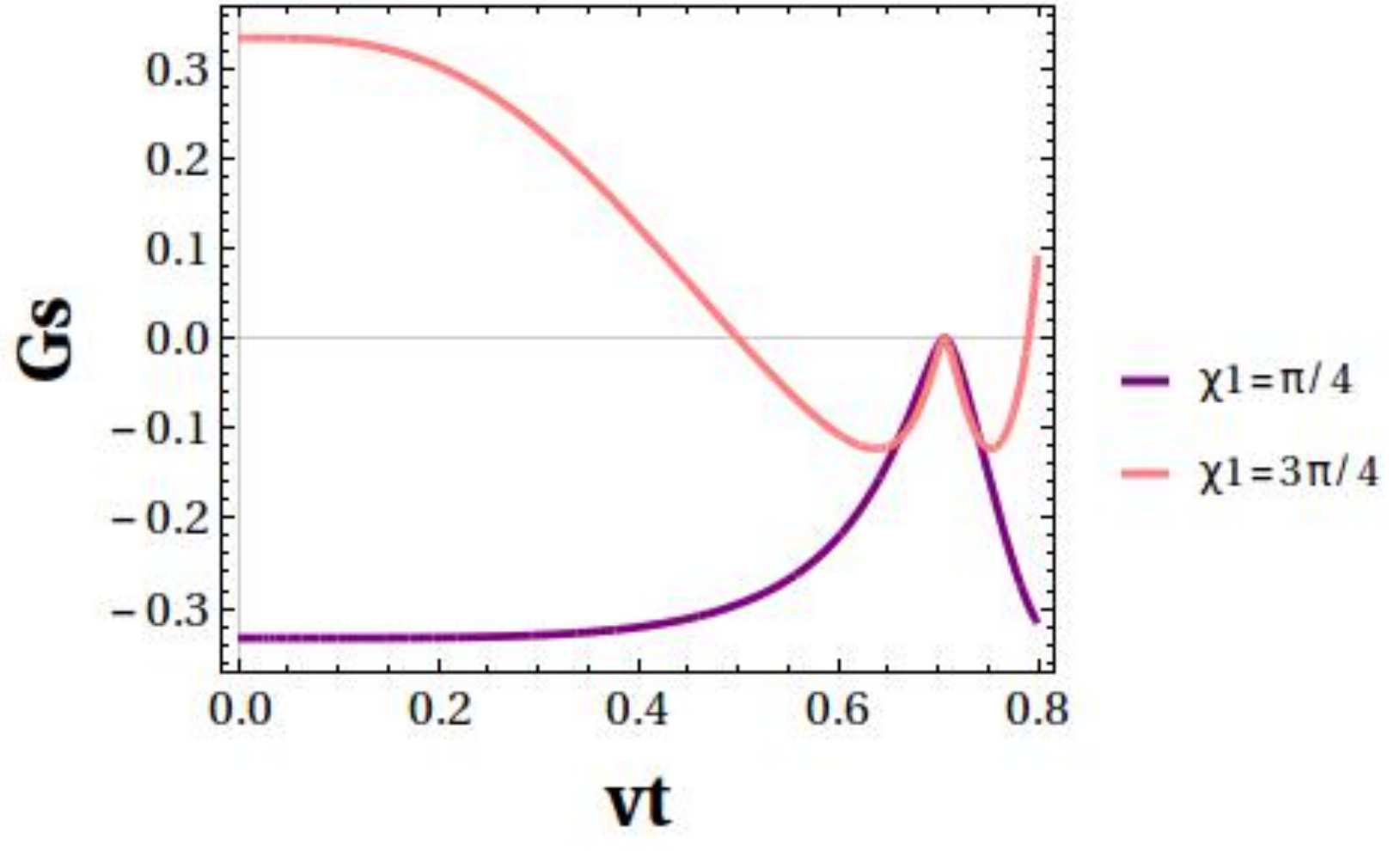}
\caption{Spin conductance ($G_s$) as a function of tilt parameter $v_t$. We fix the parameter $\chi_2=\pi/4$ and $k_F D=0$. In the upper plot, we choose the value of $\chi_1$ is $\pi/2$ and $\pi$. The spin conductance has zero value at these value of $\chi_1$ in untilted case. In the lower plot, we choose the value of $\chi_1$ is $\pi/4$ and $3\pi/4$. The spin conductance has maximum magnitude at these value of $\chi_1$ in untilted case.}
\label{spin-tilt-low}
\end{figure}
\begin{figure}
\center
\includegraphics[width=2.0in]{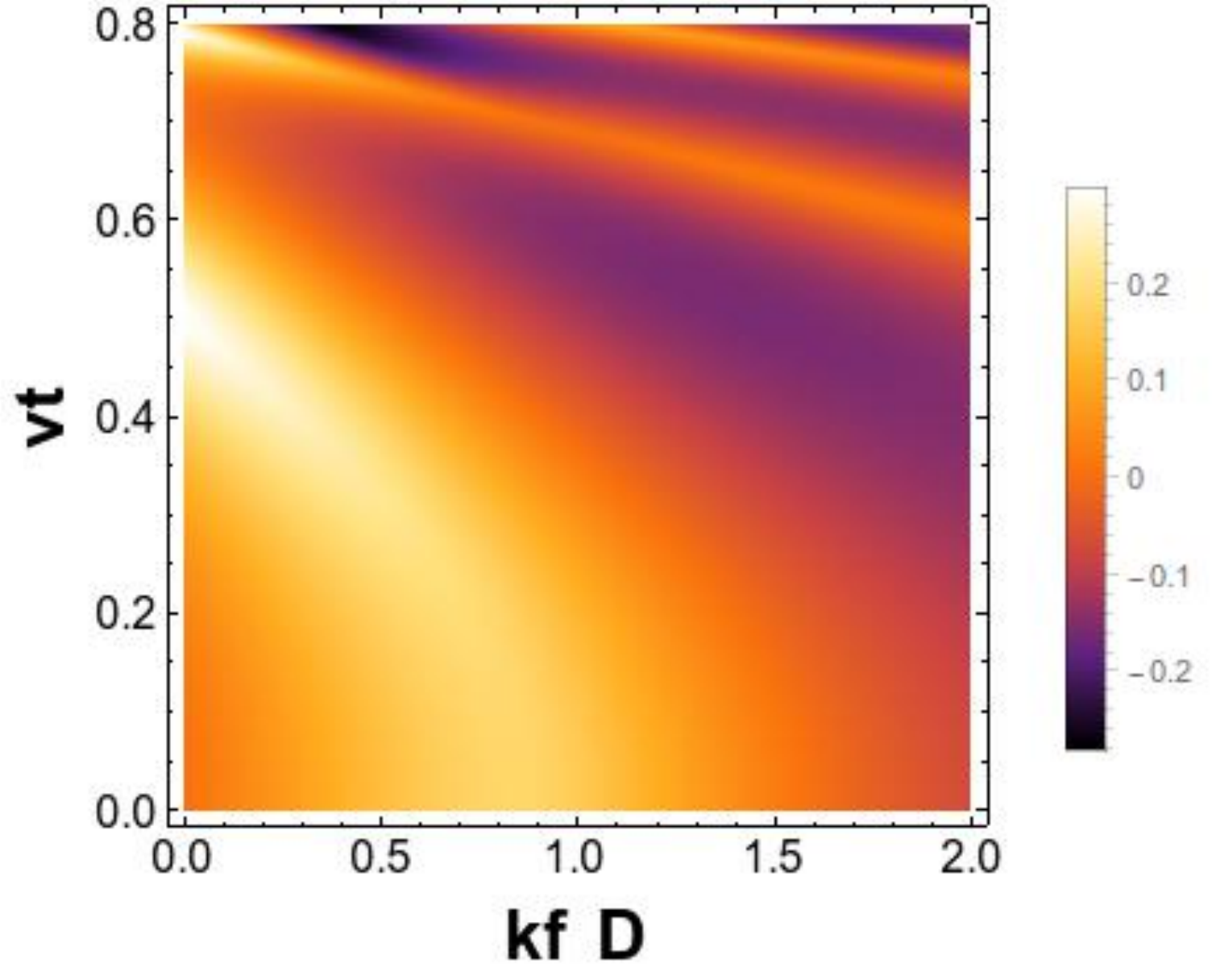}
\caption{Density plot of spin conductance ($G_s$) as a function of $v_t$ and $k_F D$. The other parameters are $\chi_2=\pi/4$ and $\chi_1=\pi/2$.}
\label{fermi-tilt}
\end{figure}
For finite tilt in the Dirac cone, the quantum interference condition discussed above gets modified due to the finite value of $v_t$ in Eq.(\ref{sol-part}). The transmission coefficient $T_\sigma$ has a maximum ($=1$) and minimum value at $\chi_{\sigma}=n\pi (1-v^2_t)$ and $(2n+1)\pi (1-v^2_t)/2$ respectively. The periodicity of spin conductances $G_{\uparrow,\downarrow}$ is now $\pi(1-v^2_t)$ shown in Fig.(\ref{spin-cond-ud}) and in Fig.(\ref{spin-cond}). The phase difference between $G_{\uparrow}$ and $G_{\downarrow}$ is now $2Jk_FD/E_F(1-v^2_t)$ and we expect a large (i.e., $\pm 1/3$) spin conductance $G_s$ at $\chi_2=(2n+1)(1-v_t^2)^2\pi/4$ (i.e. at $J/E_F=(2n+1)\pi(1-v^2_t)^2/4k_F D$). One can use the expression to measure the possible value of tilt in the Dirac cone. It is clear that for a fixed value of gate voltage $\chi_1$, proximity strength $\chi_2$ and Fermi energy $k_{F}D$, the phase of the spin conductance can itself be tuned by $v_t$. Particularly, our interest is on the behavior of spin conductance at fixed values of $\chi_1$ and $\chi_2$, where it takes zero and maximum ($\pm 1/3$) values for $v_t=0$, with tilt. In the upper plot of Fig.(\ref{spin-tilt-low}), we choose $\chi_2=\pi/4$ and plot the spin conductance ($G_s$) with tilt ($v_t$) for $\chi_1=\pi/2$ and $\pi$ respectively. Note that, the spin conductance for untilted Dirac cone has zero value at these values of $\chi_1$ and $\chi_2$ (see Fig.(\ref{spin-cond})). It is shown that with a smooth variation of tilt the spin conductance can be tuned from zero to its maximum value ($\pm 1/3$). In the lower plot of Fig.(\ref{spin-tilt-low}), we choose $\chi_2=\pi/4$ and plot the spin conductance ($G_s$) with tilt ($v_t$) for $\chi_1=\pi/4$ and $3\pi/4$ respectively. The spin conductance for untilted Dirac cone has maximum magnitude at these values of $\chi_1$ and $\chi_2$ (see Fig.(\ref{spin-cond})). The spin conductance can be tuned from its maximum values to zero with a smooth variation of tilt. In Fig(\ref{spin-tilt-low}) it is also shown that a pure spin current reversal is possible with tilt even for zero and maximum values (for $v_t=0$) of spin conductance. However, the tuning of spin conductance and its sign changes with tilt can occur with any others suitable choice of gate voltage ($\chi_1$), proximity strength ($\chi_2$) and Fermi energy ($E_F$) (not shown). The observed phenomena can be explained as follows. The finite value of tilt modifies the barrier strength and proximity coupling (see Eq(\ref{trans})). An electron scattered from the ferromagnetic region experiences a different potential barrier for the spin up and spin down and hence acquire a phase shift\cite{Yoko-08,Take-PRB 11}. With the smooth variation of tilt the acquire phase shift between $G_{\uparrow}$ and $G_{\downarrow}$ can be in or out of phase which steers the spin current reversal.

Figure(\ref{fermi-tilt}) displays density plot of $G_s$ as a function of $v_t$ and $k_F D$ for a fixed value of gate voltage $\chi_1=\pi/2$ and proximity strength $\chi_2=\pi/4$. It shows that the spin current reversal can occur with variation of tilt ($v_t$) and Fermi energy ($E_F$). We check that the charge conductance $G_{\uparrow}+G_{\downarrow}$ is positive in the entire range of $v_t$ in Fig.(\ref{spin-tilt-low}) and in Fig(\ref{fermi-tilt}). So, the spin current reversal occur here is not accompanied by charge current reversal.

\begin{figure}
\center
\includegraphics[width=1.67in]{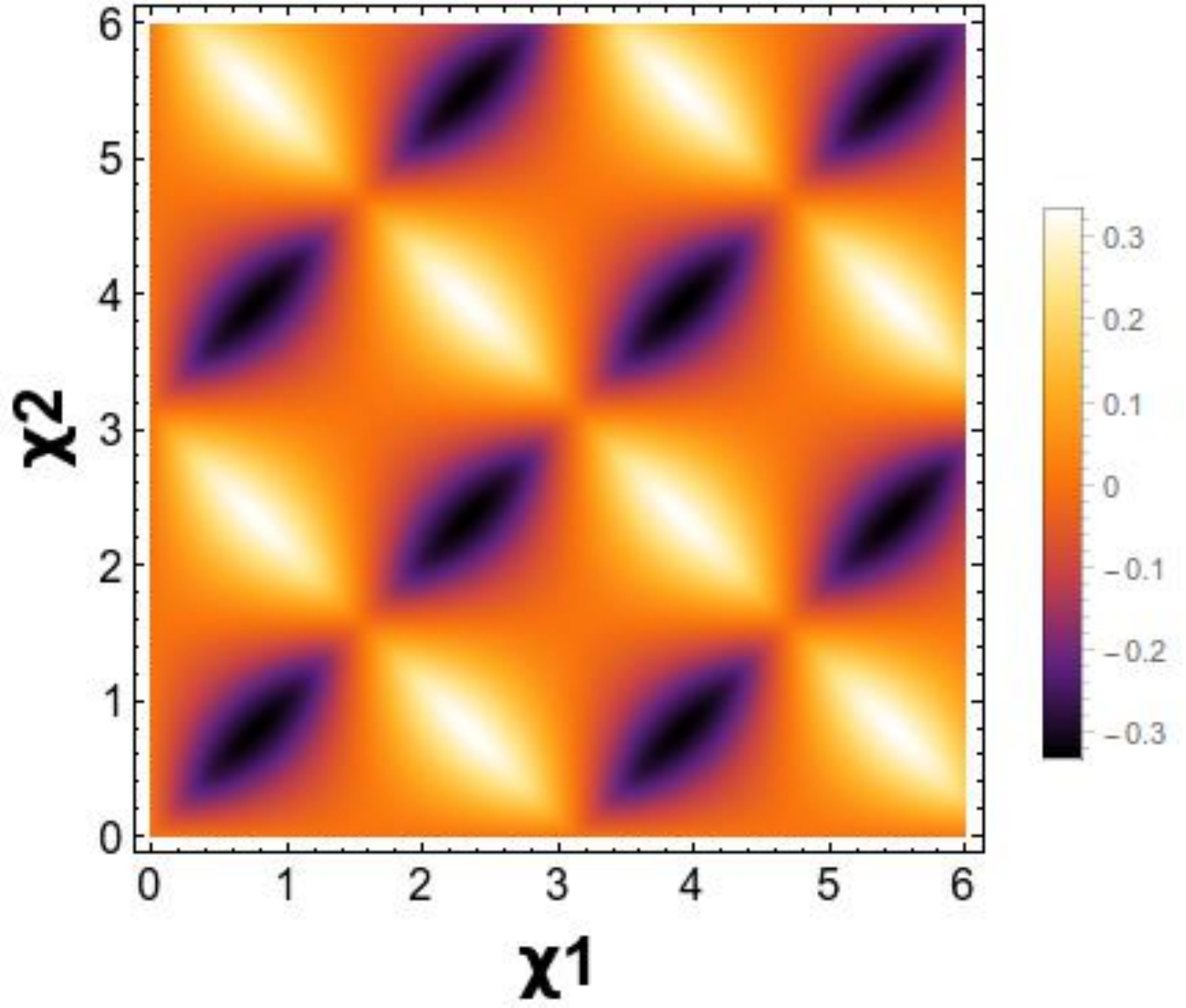}
\includegraphics[width=1.67in]{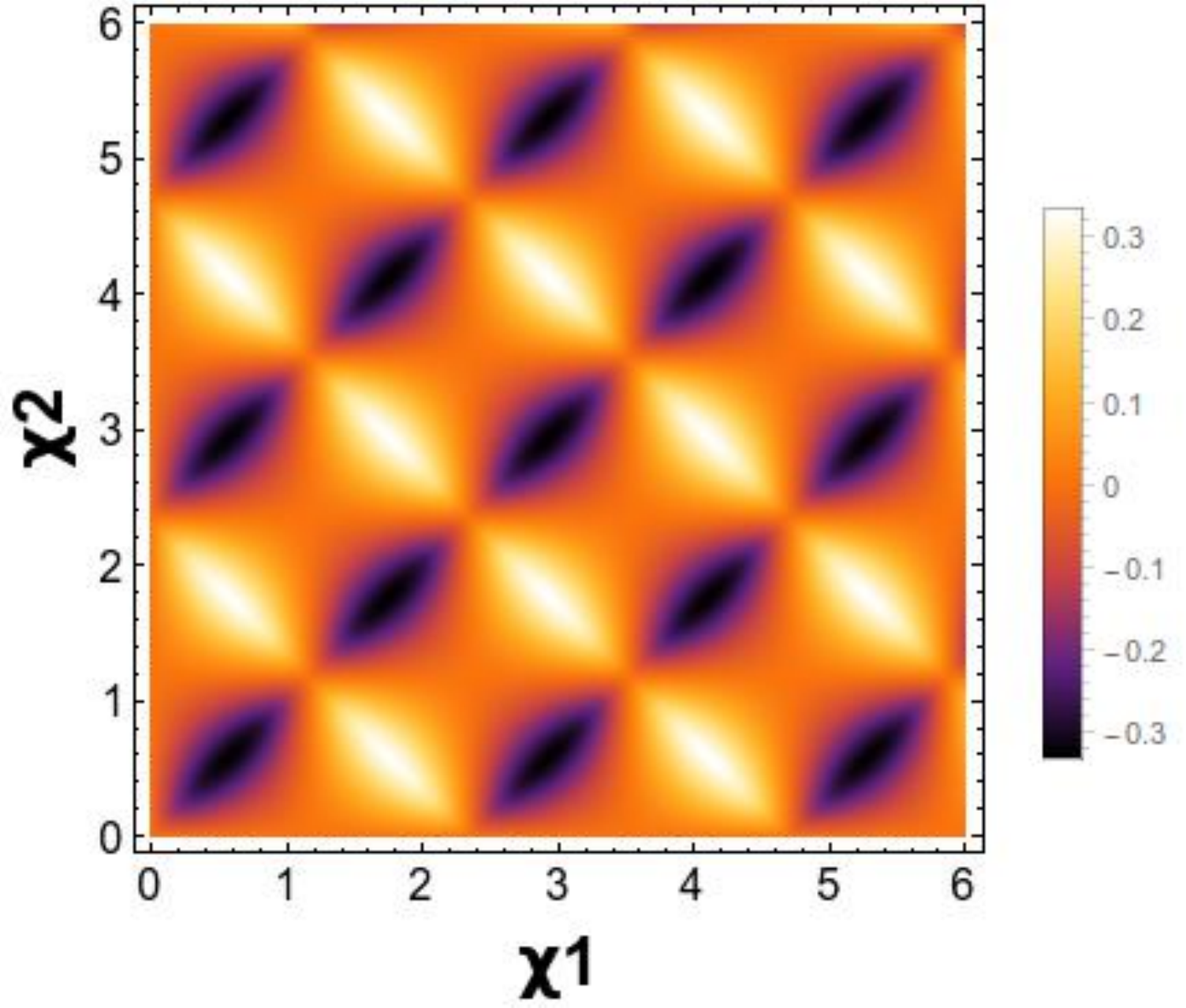}
\caption{Density plot of spin conductance ($G_s$) as a function of $\chi_1$ and $\chi_2$ for $v_t=0$ and $v_t=0.5$ respectively. We fix $k_{F}D=0.01$.}
\label{chi-nor-tilt}
\end{figure}

Figure(\ref{chi-nor-tilt}) displays density plot of $G_s$ as a function of $\chi_1$ and $\chi_2$ without tilt ($v_t=0$) and with tilt ($v_t=0.5$). $G_s$ has a $\pi$ periodicity with $\chi_2$ for $v_t=0$ (Here we choose $k_F D=0.01$). However for finite tilt, the $\pi$ periodicity is broken and it depends on $v_t$ as we discussed before. In the next section we discuss the effect of tilt on the quantum spin pumping current.

\section{Quantum Spin Pump: Effect of tilt}
The model setup for spin pumping consists of ferromagnetic region sandwich between two normal parts which act as reservoirs. We consider tilted Dirac cone in the ferromagnetic part as before. An additional metallic gate voltage is attached to the ferromagnetic part to tune the Fermi level of the electron relative to the normal regions. When magnetization vector $\vec{m}(t)$ in ferromagnetic region starts precessing (under influence of an applied magnetic field) with an adiabatic frequency $\omega$, a pure spin current (both ac and dc components) can be produced without any charge current. This pump current depends on the complex-valued parameter $g=g_r +ig_i$ and is given by,\cite{Ino-PRB 16,Tse-PRB 02}
\begin{eqnarray}
I^{pump}_{s}=\frac{\hbar}{4\pi}(g_{r}\vec{m}\times \frac{d\vec{m}}{dt}-g_{i}\frac{d\vec{m}}{dt})
\end{eqnarray}
where $g$ depends on the scattering matrix of the ferromagnetic film and is expressed as, 
\begin{eqnarray}
g=\sum_{nn'}[\delta_{nn'}-r^{\uparrow}_{nn'}(r^{\downarrow}_{nn'})^*]-t^{'\uparrow}_{nn'}(t^{'\downarrow}_{nn'})^*
\label{comp-cond}
\end{eqnarray}
where $r^{\uparrow}_{nn'}(r^{\downarrow}_{nn'})$ is a reflection coefficient for spin up (down) electrons on the normal side and $t^{'\uparrow}_{nn'}(t^{'\downarrow}_{nn'})$ is a transmission coefficient for spin up (down) electrons incident on the ferromagnetic region from the opposite reservoir\cite{Tse-PRB 02}. In absence of spin-orbit coupling, all matrix are diagonal and Eq.(\ref{comp-cond}) simplifies to \cite{Ino-PRB 16},
\begin{eqnarray}
g=g_{0}\int^{\pi/2}_{-\pi/2} d\phi \cos\phi(1-r^{\uparrow}(\phi)[r^{\downarrow}(\phi)]^*-t'^{\uparrow}(\phi)[t'^{\downarrow}(\phi)]^*)
\label{g-cond}
\end{eqnarray}
with $g_0=2k_F W/\pi$ (taking both $K$ and $K'$ points). Here, we consider that the magnetization vector rotates in $x-z$ plane and expressed as $\vec{m}(t)=m(\hat{x} \sin \omega t +\hat{z} \cos \omega t)$ (applied magnetic field along $\hat{y}$ direction). The instantaneous spin current pumped into the normal regions,
\begin{eqnarray}
I^{pump}_s(t)=\frac{\hbar\omega}{4\pi}(m^2g_r \hat{y}-m \cos \omega t g_i \hat{x}+m\sin \omega t g_{i}\hat{z})
\end{eqnarray}
time average becomes
\begin{eqnarray}
J_{s}=\frac{\omega}{2\pi}\int^{\frac{2\pi}{\omega}}_0 dt I^{pump}_{s}(t)=\frac{\hbar \omega}{4\pi}m^2 g_{r}\hat{y}
\label{pump-current}
\end{eqnarray}
and spin current per unit width is $j_s=J_s/W$. Here $g_r$ is the real part of $g$ in Eq.(\ref{g-cond}). The reflection, transmission coefficient and hence the spin conductance depend on the tilt as we discussed in the earlier section. Here we focus on the effect of tilt on the spin mixing conductance in Eq.(\ref{g-cond}) and as a result of spin pumping current. Our focus is on the spin pumping of a weakly magnetized electron gas with slightly different Fermi surface for up and down spins. The spin coherence length $\lambda=\pi/(k^{\uparrow}_F-k^{\downarrow}_F)$ in general is large for graphene-like materials in contrast to metallic ferromagnet. Therefore, the spin mixing transmission can't be neglected in Eq.(\ref{g-cond}).

The transmission coefficients is given in Eq.(\ref{trans}) and reflection coefficient reads,
\begin{eqnarray}
r^{\sigma}_k=\frac{ie^{i\phi}\sin \theta [\gamma \gamma^{\sigma}_1(e^{i\phi}+\mathcal{P}_1\mathcal{P}_2 e^{-i\phi})-(\mathcal{P}_1+\mathcal{P}_2)]}{X^{\sigma}_k}
\end{eqnarray}
Now given $r^{\sigma}_k$ and $t^{k}_\sigma$ we calculate the spin mixing conductance Eq.(\ref{g-cond}) and spin pumping current in Eq(\ref{pump-current}). 
\begin{figure}
\center
\includegraphics[width=1.65in]{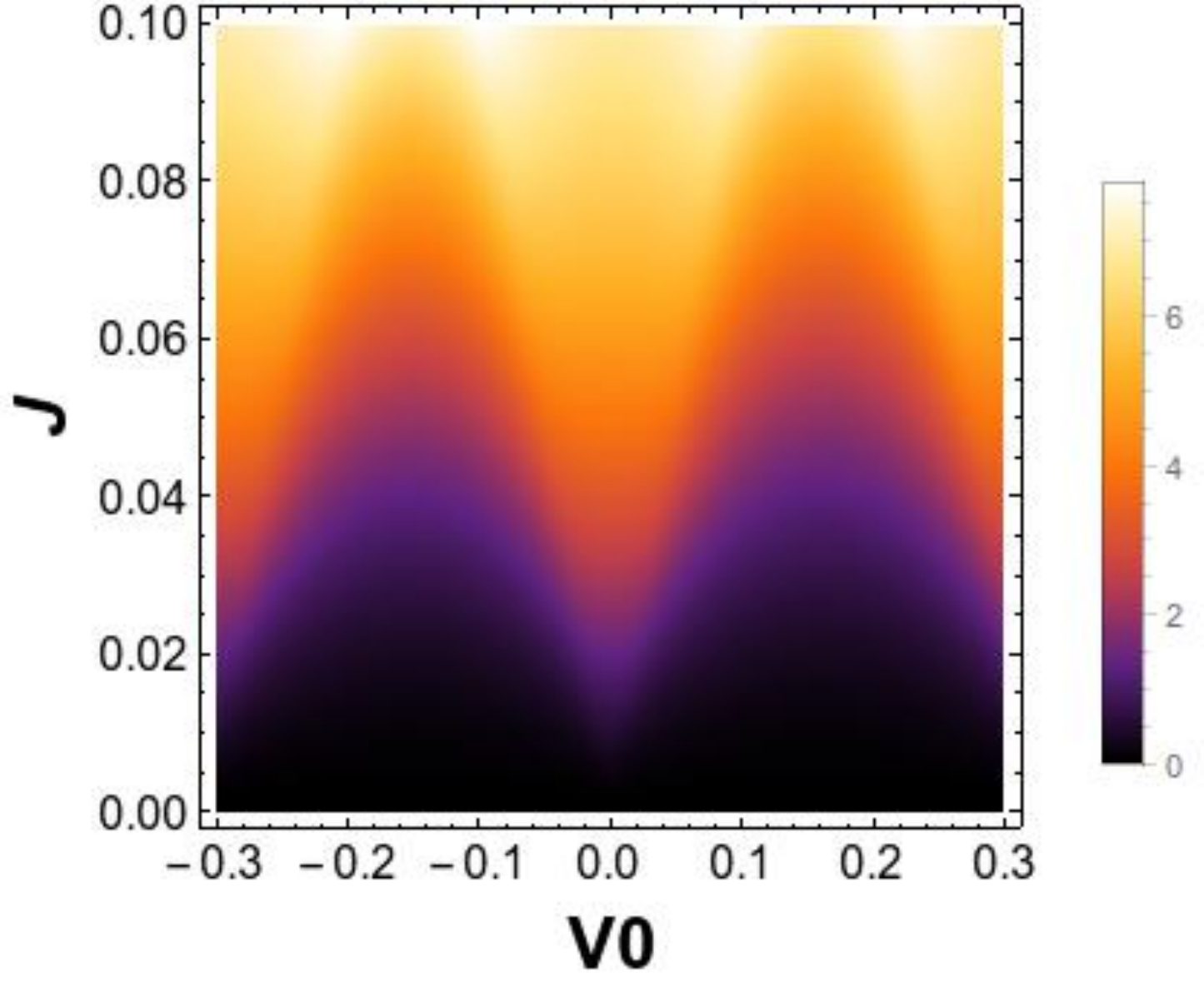}
\includegraphics[width=1.65in]{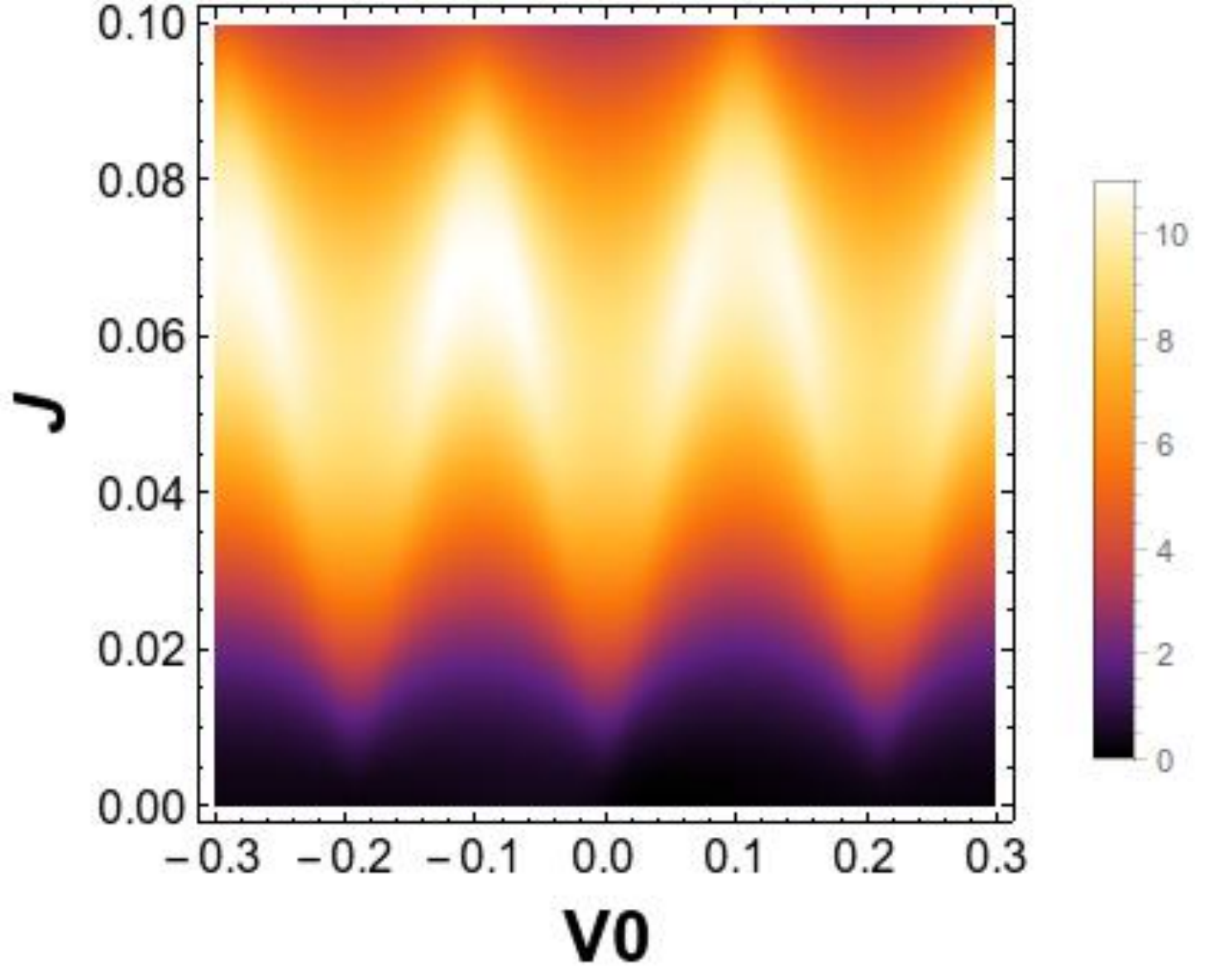}
\caption{Spin current per unit width pumped into the normal region as a function of the gate voltage $V_0$ and proximity coupling $J$. The value of tilted parameter $v_t$ is 0 in (a) and 0.6 in (b). The other parameters are $D=10$ nm, $E_F=10$ meV.}
\label{prox-gate}
\end{figure}
\begin{figure}
\center
\vskip 1cm
\includegraphics[width=2.in]{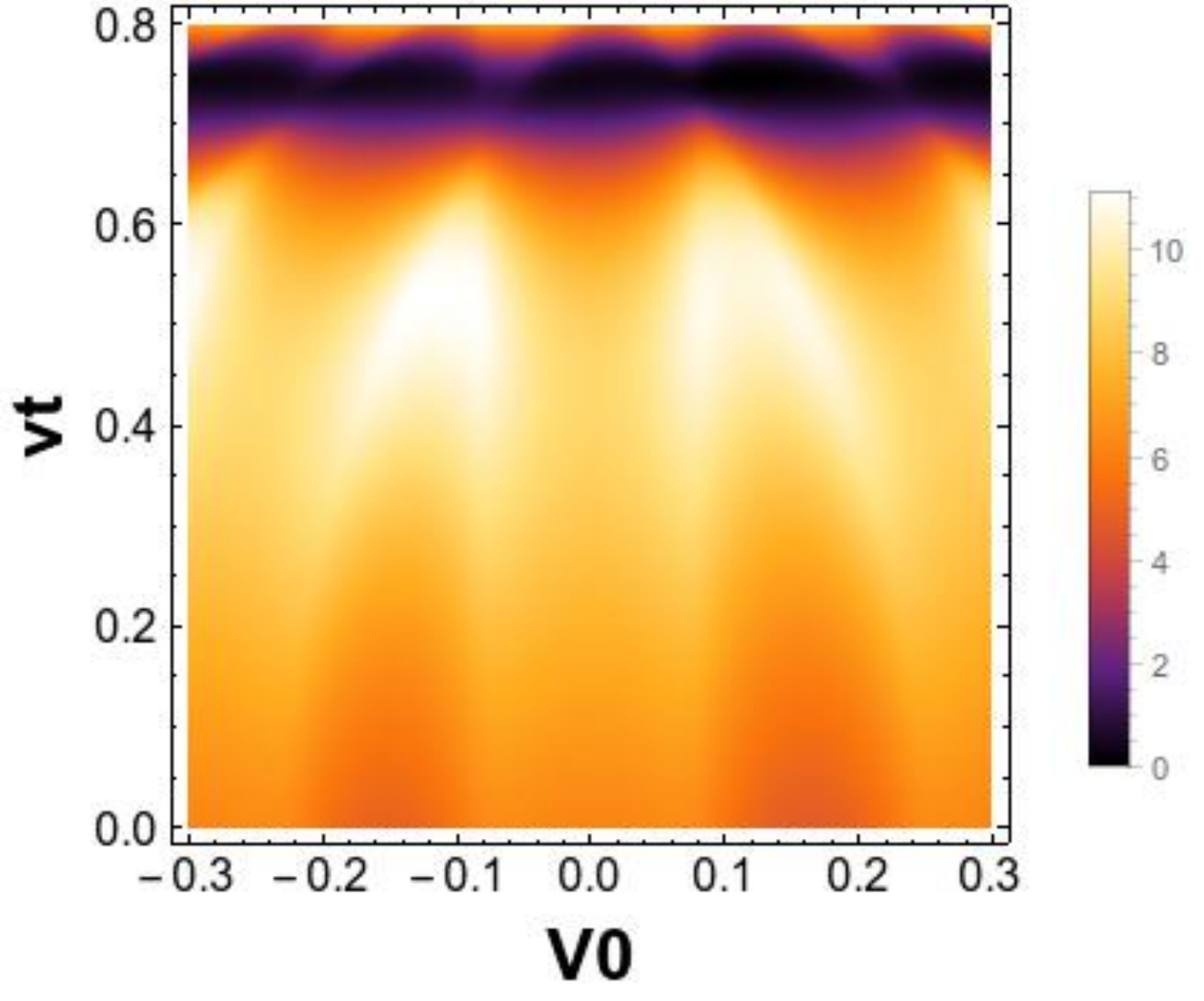}
\caption{Spin current per unit width pumped by ferromagnetic top layer as Figure(\ref{tilted-gate}) shows the density plot of spin pumping current as a function of tilt ($v_t$) and gate voltage ($V_0$).a function of tilted parameter $v_t$ and gate voltage $V_0$. The length of ferromagnetic region is $D=10$nm, $E_{F}=10$meV and $J=0.08$ eV.}
\label{tilted-gate}
\end{figure}

Figure(\ref{prox-gate}) shows the density plot of spin pumping current as a function of proximity strength ($J$) and gate voltage ($V_0$) in absence of tilt ($v_t=0$) and a finite tilt value ($v_t=0.6$). For small splitting (i.e, when the proximity strength $J$ is sufficiently low) we expect the magnitude of $I_s$ is negligibly small. A larger value of $J$ increases the spin pumping current. We choose a narrow gate width ($D=10$nm) such that the spin coherence length $\lambda$ is comparable to the length of the ferromagnet. The finite tilt enhances the spin pumping current even for a small value of proximity strength $J$.

Figure(\ref{tilted-gate}) shows the density plot of spin pumping current as a function of tilt ($v_t$) and gate voltage ($V_0$). Here we choose $J=0.08$eV ($i.e.,$ $\lambda\simeq 13$ nm). As the quantum interference problem get modified in N-F-N junction in presence of tilt, the pumping current may become zero for a finite $v_t$. Thus the tilt can lead to a spin valve effect in spin pumping experiment.

\section{Conclusions}
In summary, we have studied the effect of tilt in the Dirac cone on spin conductance and pump spin current in normal/ferromagnetic/normal graphene-like material junctions. The periodicity of oscillation of spin conductance and pump spin current are modified in presence of tilt in the Dirac cone. As result, the pure spin current reversal by tuning the tilt is possible. We also find that the spin pumping current can get enhanced with tilt even for small proximity strength. The spin mixing conductance and resulting spin pump current may become zero with the smooth variation of tilt which leads to a spin valve effect.

We must point out that our focus here is on the ballistic transport regime, although, we expect that the obtained results remain qualitatively the same in the diffusive region of transport. Our findings are valid as long as the continuum Dirac equation is valid. This requires a wide nanoribbon for the experimental realization. The tilt of the Dirac cone can be tuned by applying a pressure in a wide class of materials having no center of spatial inversion symmetry\cite{Hills-PRB 17}. Thus it could open a new route to control the spin current by tilt. Hope, our findings can be verified experimentally in near future. Although, here we have discussed only the case $v_t<1$ but we can generalize this model in type-II case ($v_t>1$) also. The future study will be in that directions.

\section{Acknowledgement} I greatly acknowledge Prof. L. E .F. Foa Torres and Prof. K. Sengupta for many useful suggestions and comments.

\end{document}